\documentstyle[11pt,aaspp4]{article}

\def\go{\mathrel{\raise.3ex\hbox{$>$}\mkern-14mu\lower0.6ex\hbox{$\sim$}}}
\def\lo{\mathrel{\raise.3ex\hbox{$<$}\mkern-14mu\lower0.6ex\hbox{$\sim$}}}
\def\be{\begin{equation}}
\def\ee{\end{equation}}
\def\hato{{\hat{\bf\Omega}}}
\def\hatb{{\hat{\bf B}}}
\def\abs{{({\rm abs})}}
\def\sc{{({\rm sc})}}
\def\eps{{\epsilon}}
\def\in{{\rm in}}
\def\out{{\rm out}}
\def\br{{\bf r}}
\def\bF{{\bf F}}
\def\bS{{\bf S}}
\def\nue{{\nu_e}}
\def\bnue{{\bar\nu_e}}
\def\numu{{\nu_\mu}}

\begin{document}
\title{
Parity Violation in Neutrino Transport and the Origin of Pulsar Kicks}

\author{Dong Lai}
\affil{
Theoretical Astrophysics, California Institute of Technology,
Pasadena, CA 91125;\\
Department of Astronomy, Cornell University, Ithaca, NY 14853\\
E-mail: dong@spacenet.tn.cornell.edu}
\and
\author{Yong-Zhong Qian}
\affil{Physics Department, 161-33, California Institute of Technology,
Pasadena, CA 91125}

\begin{abstract}
In proto-neutron stars with strong magnetic fields, the neutrino-nucleon
scattering/absorption cross sections depend on the direction
of neutrino momentum with respect to the magnetic field axis, 
a manifestation of parity violation in weak interactions. 
We study the deleptonization and thermal cooling (via neutrino 
emission) of proto-neutron stars in the presence of such asymmetric
neutrino opacities. Significant asymmetry in neutrino emission is obtained
due to multiple neutrino-nucleon scatterings.
For an ordered magnetic field threading the neutron star interior, 
the fractional asymmetry in neutrino emission is about $0.006\,(B/10^{14}
\,{\rm G})$, corresponding to a pulsar kick velocity of about $200
\,(B/10^{14}\,{\rm G})$\,km\,s$^{-1}$ for a total radiated neutrino energy of
$3\times 10^{53}$ erg. 
\end{abstract}


\section{Introduction}

Recent analyses of pulsar proper motion
(Lyne \& Lorimer 1994; Lorimer et al.~1997; Hansen \& Phinney 1997;
Cordes \& Chernoff 1998), observation of pulsar bow shocks 
(Cordes et al.~1993), and studies of pulsar-supernova remnant associations
(Frail et al.~1994) indicate that supernova explosions are asymmetric and the
neutron stars receive large kick velocities at birth ($250$--$500$ km\,s$^{-1}$
on average, but possibly with a significant population having velocities
greater than $1000$ km\,s$^{-1}$). Compelling evidence for supernova asymmetry 
also comes from the detection of geodetic precession in PSR 1913+16
(Cordes et al.~1990) and orbital plane precession in the PSR
J0045-7319/B star binary (Kaspi et al.~1996).
In addition, evolutionary considerations of
neutron star binary population also imply the existence of pulsar
kicks (e.g., Fryer et al.~1998).

Two classes of mechanisms for the natal kicks of pulsars 
have been suggested. The first class relies on 
local hydrodynamic instabilities in the proto-neutron stars
(e.g., Burrows et al.~1995; Burrows \& Hayes 1996; Janka \& M\"uller 1996) 
as well as global asymmetric perturbations seeded in the presupernova
cores (Goldreich et al.~1996).
The asymmetries in matter and temperature distributions lead to
asymmetric explosion and/or asymmetric neutrino emission, 
although the magnitude of the resulting kick velocity is still unclear.

In this {\it letter}, we focus on the second class of models in which the large
pulsar velocities arise from asymmetric neutrino emission induced by 
strong magnetic fields. Although a number of papers have been written on this
subject (Chugai 1984; Dorofeev et al.~1985; Binovatyi-Kogan 1993; Vilenkin
1995; Horowitz \& Piekarewicz 1997; Horowitz \& Li 1997), 
they are unsatisfactory for a number of
reasons: they either failed to identify the most relevant neutrino 
emission/interaction
processes or the relevant physical conditions in proto-neutron stars, or 
stopped at estimating the magnetic field effects on neutrino opacities, 
and none of them treated neutrino transport in any systematic manner. 
As a result, their estimates of the kick velocity that may
be produced are unreliable. We have carried out systematic studies of neutrino
transport in strong magnetic fields (Lai \& Qian 1998a,b) in order to pin down
the magnitude of the magnetic field effects on neutrino transport and pulsar
kicks. Here we report our most important finding on the macroscopic
consequence of asymmetric neutrino opacities due to parity violation in weak
interactions. 
More details can be found in our forthcoming
publication (Lai \& Qian 1998a, hereafter LQ).

\section{Parity Violation and Asymmetric Neutrino Opacities}

In the presence of an external magnetic field,
the differential opacity 
for neutrino-nucleon scattering, 
$\nu_{(e,\mu,\tau)}+N\rightarrow\nu_{(e,\mu,\tau)}+N$
can be written in the form:
\be
\left({d\tilde\kappa_E^\sc\over d\Omega'}\right)_{\hato\rightarrow\hato'}\!
={1\over 4\pi}\kappa_E^\sc\Bigl[1+\eps_\in\hato\cdot\hatb
+\eps_\out\hato'\cdot\hatb+{\rm const}\times(\hato\cdot\hato')\Bigr],
\label{scat}
\ee
where $\hato$ and $\hato'$ are the unit momentum vectors for 
the incoming and outgoing neutrinos, respectively,
$\hatb$ is the unit vector along the magnetic 
field direction, and $\kappa_E^\sc$ is the total scattering opacity
without magnetic field (the subscript ``E'' implies that the opacity 
depends on the neutrino energy $E$). 
The presence of the $\hatb$-depedent terms in Eq.~(\ref{scat}) 
implies that there is a
preferred direction in the opacity. This is a manifestation of parity
violation in weak interactions. 
The coefficients $\eps_{\in}$ and $\eps_{\out}$ are related to the nucleon
polarization $P$ by
\be
\eps_{\in}=2P{c_A(c_A-c_V)\over c_V^2+3c_A^2},~~
\eps_{\out}=-2P{c_A(c_A+c_V)\over c_V^2+3c_A^2},
\ee
where $c_V=-1/2,~c_A=-1.15/2$ for neutron and
$c_V=1/2-2\sin^2\theta_W=0.035,~c_A=1.37/2$ for proton 
(Raffelt \& Seckel 1995). For nondegenerate nucleons, 
the spin polarization $P$ is 
$P={\mu_m B/kT}=3.15\times 10^{-5}g_mB_{14}
\left({10~{\rm MeV}/T}\right)$,
where $\mu_m$ is the nucleon magnetic moment ($g_m=-1.913$ for neutron and
$2.793$ for proton), and $B_{14}$ is the field strength in units of
$10^{14}$ G. In the degenerate regime, we have
$P={3\mu_m B/(2\mu_N)}=4.73\times 10^{-5}g_mB_{14}\left({10~{\rm MeV}
/\mu_N}\right)$,
where $\mu_N$ is the nucleon chemical potential (or Fermi energy).
A general expression for $P$ is given in LQ.
In proto-neutron stars, one should average Eq.~(\ref{scat}) over 
composition (neutrons and protons). It is also important to note 
that the differential opacity for $\bar\nu_{e,\mu,\tau}$ scattering
on nucleons is obtained by interchanging $\eps_\in$ and
$\eps_\out$ in Eq.~(\ref{scat}) as a result of the crossing 
symmetry of the leading order matrix elements.  

The neutrino absorption opacity is also asymmetric, i.e.,
$\tilde\kappa_E^\abs=\kappa_E^\abs(1+\eps_a\hato\cdot\hatb)$,
with $\eps_a\sim P$. However, in the bulk interior
of the neutron star, this asymmetry is exactly canceled by the asymmetry
in emission (see \S4). One also expects asymmetry 
in $\nu+e\rightarrow\nu+e$. 
However, for relativistic degenerate electrons, the $\nu$-$e$ opacity
is much smaller than the $\nu$-$N$ opacity, while the electron polarization 
$P_e\simeq 2\times 10^{-5}B_{14}(200\,{\rm MeV}/\mu_e)^2$ is of the same
order as $P$ for nucleons. Therefore the contribution from 
$\nu$-$e$ scattering to asymmetric neutrino transport is small compared with
that from $\nu$-$N$ scattering.

\section{Toy Problem: Effect of Multiple Scatterings}

The asymmetry coefficients ($\eps_\in$ and $\eps_\out$)
in the opacity are extremely small even at $B_{14}=10$.
However, cumulative effect due to multiple scatterings can enhance the 
asymmetry in neutrino emission. 
To demonstrate this effect, we consider a simple toy problem in which
neutrinos leak out of a slab (with infinite extent perpendicular to
the $z$-axis which coincides with the magnetic axis) 
via biased diffusion, i.e., 
the neutrino scattering probability has the form
$\propto (1+\eps\cos\theta)$, with $\eps\ll1$ and 
$\theta$ being the angle between the $z$-axis and the 
direction of neutrino propagation after each scattering. 
The evolution of the neutrino number density $n(z,t)$ is governed by 
the Fokker-Planck equation:
\be
{\partial n\over\partial t}={1\over 3}\,c\lambda\,\nabla^2n-
{1\over 3}\, c\,\nabla\cdot (\eps n\hatb),
\label{diff}
\ee
where $\lambda$ is the (constant) mean free path of the neutrino. 
The surfaces of the slab are located at $z=\pm\lambda\tau_s$ (so
$2\tau_s$ is the total optical depth of the slab). Suppose
we start with a uniform distribution $n=n_0$ at $t=0$ (so that there is no
initial asymmetry). As neutrinos gradually diffuse out, the profile
$n(z,t)$ becomes increasingly asymmetric, and the flux asymmetry 
$\Delta F/F$ grows (where $F$ is the sum of the fluxes from
both surfaces and $\Delta F$ is the difference). 
For $t\go \tau_s^2(\lambda/c)$, the ``cooling'' front reaches 
the center of the slab, and $\Delta F/F$ saturates to 
constant value, $\Delta F/F\simeq \eps\tau_s/2$  (valid for
$\eps\tau_s\ll 1$). This constant value can be derived
analytically from a quasi-steady-state analysis.
The fractional asymmetry in the total neutrino radiation energy is 
$0.34\,\eps\tau_s$. 

In proto-neutron stars, the neutrino optical depth $\tau$
is large. A rough estimate can be made by setting the diffusion time
$\sim 3R^2/(\lambda c)$ ($R$ is the neutron star radius) 
to $1$\,s (the spread in the arrival time of SN 1987A neutrinos was $10$\,s),
which gives $\tau\sim 10^4$. This would imply a large enhancement in
neutrino emission asymmetry. Several complications arise when one 
applies the result of the toy problem to real neutron stars:
As shown in \S4, the asymmetry in neutrino flux is determined by
the coefficient $\eps_\out-\eps_\in$ and the local neutrino energy
distribution function. Due to the crossing symmetry, we have
$(\eps_\out-\eps_\in)_\nu=-(\eps_\out-\eps_\in)_{\bar\nu}$.
Because $\nu_{\mu(\tau)}$ and $\bar\nu_{\mu(\tau)}$ have the same local
energy distribution function, 
the ``drift fluxes'' (see Eq.~[\ref{fe}]) associated with 
$\nu_{\mu(\tau)}$ and $\bar\nu_{\mu(\tau)}$ exactly cancel.
The ``drift fluxes'' of $\nu_e$ and 
$\bar\nu_e$ do not cancel because a newly-formed neutron star
is lepton-rich, i.e., it contains more $\nu_e$ than $\bar\nu_e$,
and the typical energy of $\nu_e$ is also different from $\bar\nu_e$.
As the neutron star is deleptonized, the difference between $\nu_e$ and
$\bar\nu_e$ becomes smaller, and we expect the asymmetry in the total 
neutrino flux to gradually diminish. Another issue which is not clear 
from the toy problem concerns the role of neutrino absorption: 
for electron-type neutrinos, absorption is more important 
than scattering (by a factor of a few) in the deleptonization phase.
Does the neutrino absorption completely wipe out the
asymmetry associated with multiple scatterings? 

\section{Thermal Evolution of Proto-Neutron Stars with Asymmetric 
Neutrino Opacities}

We now describe our calculation of the thermal evolution of proto-neutron 
stars in the presence of the asymmetric neutrino opacities discussed in \S2. 
We start from the general transport equation for the
spectral intensity $I_E=I_E(\br,\hato,t)$ of a given neutrino species:
\be
{\partial I_E\over\partial t}+\hato\cdot\nabla I_E
=\rho\,\tilde\kappa_E^{\abs}(I_E^{(FD)}\!-I_E)
+\rho\,(1-f_E)\!\!\int\!\left(\!{d\tilde\kappa_E^\sc\over
d\Omega}\!\!\right)_{\!\!\!\hato'\rightarrow\hato}\!\!\!\!\!\!\!\!
I_E'\,d\Omega' -\rho\,I_E\int\!\left(\!{d\tilde\kappa_E^\sc\over
d\Omega'}\!\!\right)_{\!\!\!\hato\rightarrow\hato'}\!\!\!\!\!\!\!\!\!
(1-f_E')\,d\Omega',
\label{IE}
\ee
where $I_E'=I_E(\br,\hato',t)$, $f_E=(2\pi)^3I_E/E^3$ is the neutrino
occupation number,
and $I_E^{(FD)}$ is the Fermi-Dirac distribution function for neutrinos:
\be
I_E^{(FD)}(T)={c\over 4\pi}U_{E}^{(FD)}=
{E^3\over (2\pi)^3}{1\over e^{(E-\mu_\nu)/T}+1}
\ee
($U_{E}^{(FD)}$ is the corresponding energy density),
with matter temperature $T$ and neutrino chemical potential
$\mu_\nu$. In Eq.~(\ref{IE}) the absorption opacity $\tilde\kappa_E^\abs$
already includes the effect of stimulated absorption of neutrinos.
Multiplying Eq.~(\ref{IE}) by $\hato$ and then integrating over $\hato$,
we obtain the first order moment equation:
\be
\bF_E =-{c\over 3\,\rho\,\kappa_E^{(t)}}\nabla U_E
+\eps_a{\kappa_E^\abs\over 3\,\kappa_E^{(t)}}
c\,(U_{E}^{(FD)}-U_E)\,\hatb
+(\eps_\out-\eps_\in){\kappa_E^\sc\over 3\,\kappa_E^{(t)}}
\,c\,U_E\,(1-f_E)\,\hatb,
\label{fe}
\ee
where $U_E,\,\bF_E$ is the spectral energy density and flux,
and $\kappa_E^{(t)}=\kappa_E^\sc+\kappa_E^\abs$ is the total opacity.
The neutrino number flux is simply given by $\bS_E=\bF_E/E$.
Equation (\ref{fe})  reveals that in addition to 
the usual ``diffusive'' flux (the first term), there is a ``drift'' flux
(the $\hatb$-dependent terms) along the magnetic field direction
due to the asymmetric neutrino
opacities. Note that when $U_E=U_E^{(FD)}$ (i.e., when
neutrinos are in thermal equilibrium with matter), the asymmetric absorption
opacity does not contribute to the drift flux, since in this case asymmetric
emission exactly cancels asymmetric absorption. In the following we shall 
focus on the bulk interior of the star (below the layer where neutrinos
decouple from the matter), where $U_E$ for $\nu_e$ has the Fermi-Dirac
form with chemical potential $\mu_{\nu_e}$, while $\nu_\mu$ and $\nu_\tau$
have zero chemical potential. 

For a given neutrino species $i$, the total opacity can be written 
in the form $\kappa_{Ei}^{(t)}=\kappa_{Ei}^\abs+\kappa_{Ei}^\abs
=\kappa_i\left({E/E_0}\right)^2$
($E_0$ is a fiducial energy). In general, because of nucleon degeneracy and/or
neutrino degeneracy, there can be additional $E$-dependence 
in the prefactor $\kappa_i$ (Burrows \& Lattimer 1986). 
In these cases, we evaluate $\kappa_i$ at an appropriate mean 
neutrino energy. If we make a further assumption that $\nu_e$ and $\bar\nu_e$
have the same form of opacities (Burrows \& Lattimer 1986), 
then an analytical expression for the total energy flux (integrated over
neutrino energy) associated with $\nu_e$ and $\bar\nu_e$ can be obtained:
\be
\bF_{\nu_e}+\bF_{\bar\nu_e}=-{c E_0^2\over 6\pi^2\rho\,\kappa_\nue}\nabla
\left[T^2\left({\eta_\nue^2\over 2}+{\pi^2\over 6}\right)\right]
+{1\over 2\pi^2}\beta_\nue\eps\,c\,
T^4\left({\eta_\nue^3\over 3}+{\pi^2\eta_\nue\over 3}\right)\hatb,
\label{fluxe}\ee
where we have defined
$\eta_\nue\equiv {\mu_\nue/T}$,
$\beta_\nue\equiv{\kappa_\nue^\sc/\kappa_\nue^{(t)}}$, 
and $\eps\equiv\eps_\nue\equiv(\eps_\out-\eps_\in)_\nue$.
In deriving Eq.~(\ref{fluxe}) we have used the relation $\mu_\bnue=-\mu_\nue$
and $\eps_\bnue=-\eps_\nue$.
Similarly, the net electron lepton number flux is given by
\begin{eqnarray}
\bS_{\nu_e}-\bS_{\bar\nu_e}=-{c E_0^2\over
6\pi^2\rho\,\kappa_\nue}\nabla\Bigl (T\eta_\nue\Bigr)
+{1\over 3\pi^2}\beta_\nue\eps\,c\,
T^3\left({\eta_\nue^2\over 2}+{\pi^2\over 6}\right)\hatb.
\label{sluxe}\end{eqnarray}
We shall group $\nu_\mu,~\bar\nu_\mu,~\nu_\tau$ and $\bar\nu_\tau$
together and denote them as $\nu_\mu$. With opacity
$\kappa_E=\kappa_\numu (E/E_0)^2$, 
the total energy flux carried by these neutrinos is given by
\be
\bF_\numu=-{cE_0^2\over 18\,\rho\,\kappa_\numu}\nabla T^2.
\ee
As noted earlier, the drift flux associated with $\mu$-type neutrinos
is zero.
The equations governing the 
thermal evolution of a proto-neutron star can be written as
\begin{eqnarray}
{\partial Y_L\over\partial t}&=&
-{1\over n}\nabla\cdot (\bS_\nue-\bS_\bnue),\\
{\partial U\over\partial t}&=&-\nabla\cdot\bF=-\nabla\cdot(\bF_\nue
+\bF_\bnue+\bF_\numu).
\end{eqnarray}
Here $Y_L=n_L/n=Y_e+Y_\nu=(Y_{e^-}-Y_{e^+})+(Y_\nue-Y_\bnue)$ is the
lepton number fraction ($n$ is the baryon number density), and $U$ is
the internal energy density of the medium 
(a sum of contributions from neutrinos,
electrons, positrons, photons, and nucleons). Note that inside the star,
$\beta$-equilibrium holds to a good approximation, thus we have
$\mu_\nue+\mu_n=\mu_p+\mu_e-1.293$~MeV 
($\mu_n,\,\mu_p$ do not include rest mass). 
For simplicity, 
we have neglected gravitational contraction and 
have adopted a Newtonian treatment.

We assume a uniform magnetic field in the $z$-direction throughout the 
neutron star. When the asymmetry is small, we can expand $T(\br,t)$ and
$Y_L(\br,t)$ as
\be
T(\br,t)=T_0(r,t)+\Delta T(r,t)\cos\theta,~~~~~~~
Y_L(\br,t)=Y_{L0}(r,t)+\Delta Y_L(r,t)\cos\theta,
\ee
and similarly for other relevant quantities. The angular dependence can 
then be factored out. 
We solve the evolutionary equations numerically using an explicit 
finite difference code. 

Our model neutron star has a mass $M=1.38M_\odot$ and a radius $R=11$\,km, 
with a density profile resembling a $\Gamma=3$ polytrope (the central density 
is $\rho_c=8\times 10^{14}$ g\,cm$^{-3}$). The initial 
temperature and $Y_L$ profiles are shown in Fig.~1. 
In the stellar interior, we have $Y_e=0.28$, 
$Y_L=0.366$. The temperature is $8$ MeV in the inner core
($M\lo 0.7M_\odot$) and reaches $22$ MeV in the outer core due to shock
heating. These initial conditions are typical of those found in
core collapse simulations.
The initial chemical potentials of various particles at the center are
$\mu_n=97.7$\,MeV, $\mu_p=52$\,MeV, $\mu_e=313$\,MeV, and $\mu_\nu=266$\,MeV. 
We assume that there is no asymmetry at $t=0$. 
We adopt the zero boundary condition $T=0$ at the stellar surface. 
Since the neutrino flux asymmetry mainly accumulates in the interior 
of the star, our crude treatment of the surface boundary condition is
adequate. 

Figure 1 depicts the $T_0,\,Y_{L0}$ profiles
and the asymmetric perturbations ($\Delta T$ and $\Delta Y_L$)
as functions of time for a magnetic field of strength
$B_{14}=5$. The evolution of
the $T_0$ and $Y_{L0}$ profiles is similar to that found by
Burrow \& Lattimer (1986): The deleptonization phase ($t\lo 10$ s) is
dominated by the lepton number diffusion. In this early phase a heat wave 
moves inward, raising the core temperature.
This is followed by 
a thermal cooling phase in which the core temperature decreases
gradually due to energy diffusion. The asymmetries in the temperature
and lepton number profiles increase during the deleptonization
phase, reaching a maximum at $t\sim 8$ s (This behavior is
analogous to that found in the toy problem discussed in \S 3). 
This growth of asymmetry is also reflected by the increase in 
the ratio $|\Delta L_z/L|$ (where $L$ is the
luminosity, and $\Delta L_z/c$ is the net $z$-momentum 
carried away by the neutrinos per unit time) as shown in Fig.~2(b).
As the neutron star becomes more lepton-depleted, the difference 
(in number and energy) between $\nu_e$ and $\bar\nu_e$ diminishes, 
and the asymmetry gradually declines. 
Fig.~2(c) shows the asymmetry in the integrated neutrino luminosity
[${\cal E}(t)=\int_0^t\!dt\,L(t)$] as a function of time. The 
asymmetry in the total radiated neutrino energy is
$\alpha=|\Delta{\cal E}_z/{\cal E}|\simeq 0.006\,B_{14}$. 
This result is smaller than 
suggested by the toy problem (\S3) mainly because 
the energy carried away by electron neutrinos during the deleptonization
phase is only about $10\%$ of the total neutrino energy released.
In addition, geometric factors and cancellation of asymmetries associated 
with neutrons and protons tend to reduce the total asymmetry in 
neutrino emission. 

\section{Conclusion}

In this paper, we have identified the most efficient way of generating
natal pulsar kicks based on asymmetric neutrino emission 
induced by magnetic fields. The key point of this mechanism is parity 
violation in weak interactions. The resulting pulsar kick velocity is 
\be 
V_{\rm kick}\simeq 1000\,\left({\alpha\over 0.084}\right)
\left({E_{\rm tot}\over 10^{53}\,{\rm erg}}\right)
\left({1.4\,M_\odot\over M}\right)
\simeq 214\,\left({\langle B_z\rangle\over 10^{14}\,{\rm G}}\right)
\left({E_{\rm tot}\over 3\times 10^{53}\,{\rm erg}}\right)
\left({1.4\,M_\odot\over M}\right)\,{\rm km\,s}^{-1},
\ee
where $E_{\rm tot}$ is the total energy released by neutrinos (of
all species) from
the proto-neutron star, and $\langle B_z\rangle$ is the (averaged) 
ordered component of the magnetic field in the star.
In reality, rotation (if it is misaligned with the magnetic axis)
tends to reduce the kick velocity.
We also note that the asymmetric $\nu_e$ and $\bar\nu_e$ emission from
the proto-neutron star would give rise to asymmetric neutrino heating
behind the stalled shock in the delayed supernova mechanism.
This asymmetric heating could lead to asymmetric supernova explosions.

Another related (but distinct) kick mechanism relies on 
the fact that $\nu_e$ and $\bar\nu_e$
absorption opacities near the neutrinosphere 
depend on the local magnetic field 
strength (due to quantization of $e^-$ and $e^+$ energy levels). If the 
magnitude of magnetic field in the north pole is different from 
that in the south pole, then asymmetric neutrino flux can be generated. 
We have shown in Lai \& Qian (1998b) that this 
kick mechanism requires a much larger field strength than the mechanism
considered in this letter.

It is plausible that magnetic fields stronger than $10^{15}$\,G 
can be generated in proto-neutron stars (Thompson \& Duncan 1993), 
and several pieces of evidence (albeit tentative) have been suggested in
the literature (e.g., Vasisht \& Gotthelf 1997). 
While hydrodynamically-driven kick mechanisms may still be viable,
(depending on the magnitude of asymmetry seeded in the presupernova core),
the possibility that a significant number of
pulsars have kick velocities greater than 1000\,km\,s$^{-1}$ (e.g.,
Cordes \& Chernoff 1998) may well require large magnetic fields 
($10^{15}$\,G) to be present in proto-neutron stars.  

\acknowledgments

We thank Sterl Phinney for useful discussion at the early stage of this 
work. D.L. also thanks Ira Wasserman for discussion. This work was
started while D.L. held the Richard C. Tolman Fellowship at
Caltech. Additional support was provided by NASA grant NAG 5-2756 and
NSF grant AST-9417371. Y.Q. is supported by the David W. Morrisroe 
Fellowship at Caltech.


\bigskip

\begin{figure}
\plotone{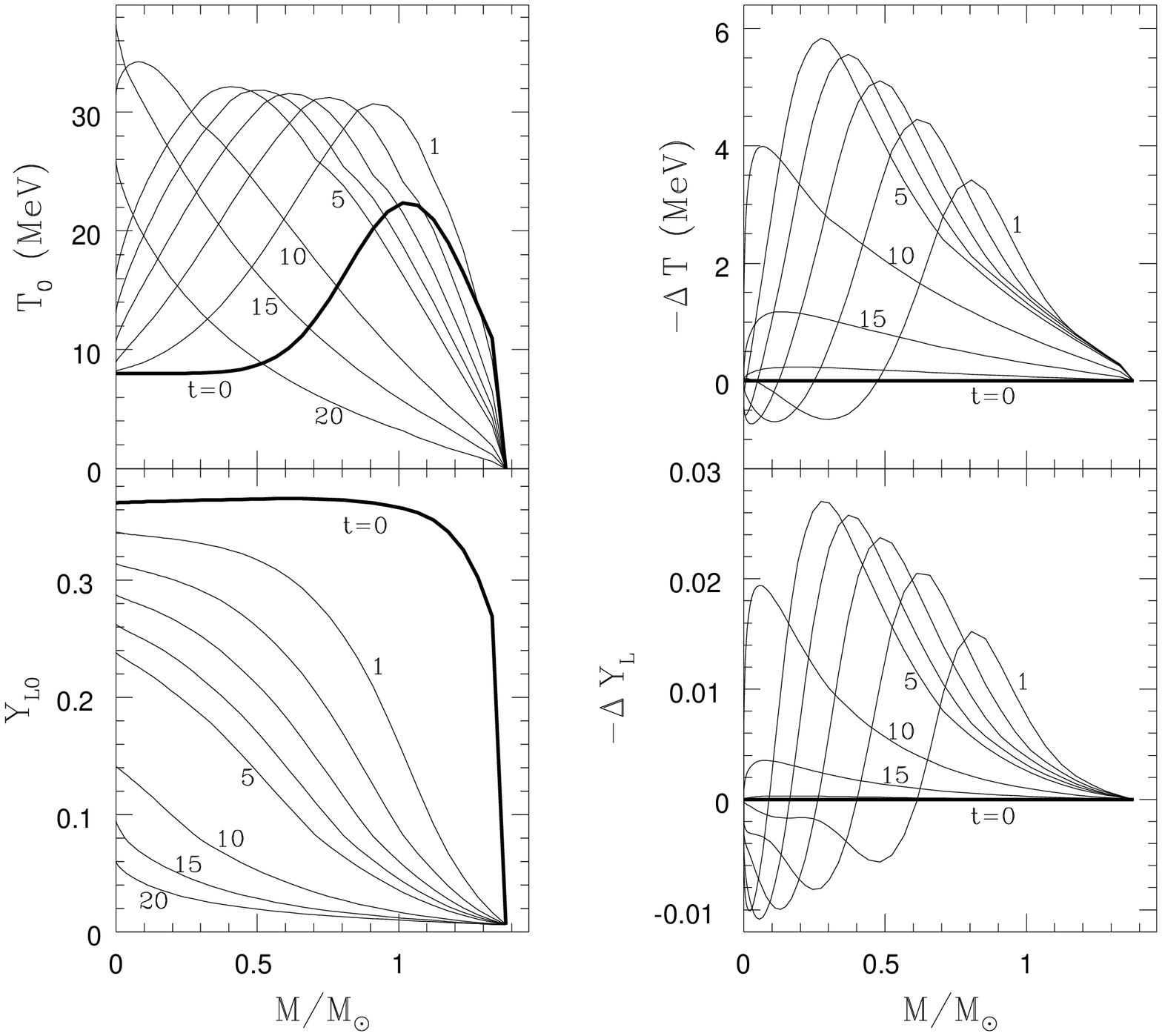}
\caption{
Thermal evolution of a proto-neutron star. 
The distributions of temperature and lepton number fraction 
(the left panels) and the asymmetric perturbations 
(the right panels) are shown as functions 
of time for
a uniform magnetic field of strength $B_{14}=5$.
}
\end{figure}

\begin{figure}
\plotone{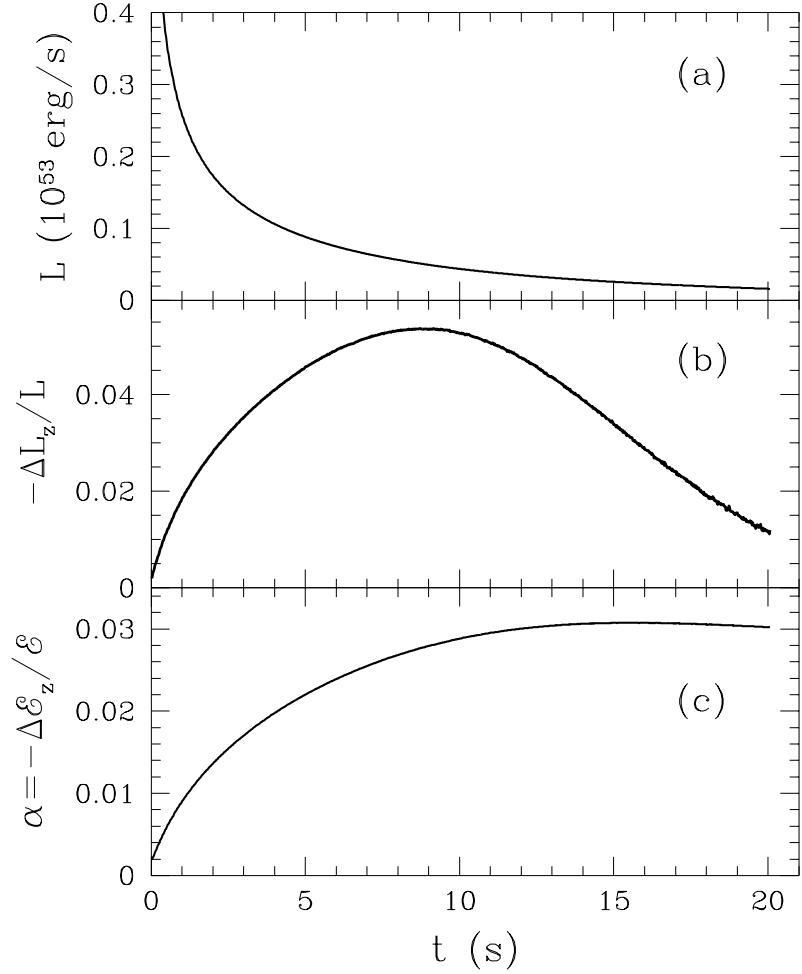}
\caption{
The time evolution of (a) total neutrino luminosity $L$, 
(b) fractional asymmetry in luminosity $\Delta L_z/L$,
and (c) fractional asymmetry in the integrated luminosity $\alpha$,
corresponding to the evolution depicted in Fig.~1.
}
\end{figure}

\end{document}